\documentclass[fleqn,twoside]{article}
\usepackage{espcrc2,colordvi,psfig,epsfig}
\newcommand{\fillrhd}{{\rhd\kern-.81em\bullet}\kern-.47em\triangleright}

\title{$N$-ality and topology at finite temperature\thanks{Presented by H. Panagopoulos}}

\author{L. Del Debbio\address{CERN, Department of Physics, TH Division, 
                                       CH-1211 Geneva 23, Switzerland},
        H. Panagopoulos\address{Department of Physics, University of
        Cyprus, Lefkosia CY-1678, Cyprus}, E. Vicari\address{Dipartimento di Fisica,
        Universit\`a di Pisa, I-57125 Pisa, Italy}}  

\def\spose#1{\hbox to 0pt{#1\hss}}
\def\ltapprox{\mathrel{\spose{\lower 3pt\hbox{$\mathchar"218$}}
 \raise 2.0pt\hbox{$\mathchar"13C$}}}
\def\gtapprox{\mathrel{\spose{\lower 3pt\hbox{$\mathchar"218$}}
 \raise 2.0pt\hbox{$\mathchar"13E$}}}

\begin{document}

\begin{abstract}
We study \cite{DPV1} the spectrum of confining strings in SU(3) pure
gauge theory, in different representations of the gauge group.  Our
results provide direct evidence that the string spectrum agrees with
predictions based on $n$-ality.  We also investigate \cite{DPV2} the
large-$N$ behavior of the topological susceptibility $\chi$ in
four-dimensional SU($N$) gauge theories at finite temperature, and in
particular across the finite-temperature transition at $T_c$. The
results indicate that $\chi$ has a nonvanishing large-$N$ limit for
$T<T_c$, as at $T=0$, and that the topological properties remain
substantially unchanged in the low-temperature phase.  On the other
hand, above the deconfinement phase transition, $\chi$ shows a large
suppression.  The comparison between the data for $N=4$ and $N=6$
hints at a vanishing large-$N$ limit for $T>T_c$.
\end{abstract}

\maketitle
\section{N-ality}
The spectrum of confining strings in $4-d$ SU($N$) gauge theories has been
much investigated recently.
Several numerical studies on the lattice have 
provided results for colour sources associated with
representations higher than the 
fundamental (see~\cite{DPV1,DPV2} for extended lists of references).
By general arguments, the string tension  must
depend only on the  
$n$-ality, $k={\rm mod}(l, N)$, of a  representation built out of
the (anti-)symmetrized tensor
product of $l$ copies of the fundamental representation.
The confining string with $n$-ality $k$ is usually
called  $k$-string, and $\sigma_k$ is its string tension.
Using charge conjugation, $\sigma_k=\sigma_{N-k}$.
As a consequence, $SU(3)$ has
only one independent string tension determining the large distance
behavior of the potential for $k\ne 0$.
One must consider larger values of $N$
to look for distinct $k$-strings.

Model-independent results predict that, in the $N\to\infty$ limit, the
$k$-string ratio ${\sigma_k/\sigma} \to k$ with corrections that are
parametrically of order $1/N^2$~\cite{Armoni:2003nz}. This excludes
the Casimir scaling law as an exact formula. Another interesting
hypothesis which has been put forward in the context of supersymmetric
theories, the sine scaling law, is consistent with this
constraint. Lattice results for $N=4,5,6$ show a nontrivial spectrum
for the $k$-strings~\cite{DPRV-02,LT-01}, which is well approximated
by the sine formula (see \cite{Strassler:1998id} and references therein). 

On the other hand, numerical results for different representations
with the same $n$-ality apparently \Blue{contradict} the picture that
$n$-ality is what really matters.  For example, MC data in SU(3) for
Wilson loops show apparently area laws up to rather large distances,
approximately 1 fm, also for representations with zero $n$-ality, and
the extracted string tensions turn out to be consistent with Casimir
scaling~\cite{Bali:2000un,Deldar:1999vi}.

In the lattice study of Ref.~\cite{LT-01,DPRV-02}, considering larger
values of $N$, the $k$-string tensions were extracted from torelon
masses, i.e. from the exponential decay of correlations of characters
of Polyakov lines.  In Ref.~\cite{DPRV-02}, while antisymmetric
representations provided rather clean measurements of $\sigma_k$
reproducing the sine formula, the numerical results for the symmetric
representations suggested different values of the corresponding string
tensions.  For example, in the case of rank 2, $\sigma_{\rm
sym}/\sigma\gtapprox 2$, which is approximately the value suggested by
Casimir scaling or by the propagation of two noninteracting
fundamental strings.

These results that apparently contradict $n$-ality may be explained by
arguing that standard colour sources, associated with representations
different from the antisymmetric ones, have very small overlap with
the stable $k$-string states, being suppressed by powers of $1/N^2$ in
the large-$N$ limit, and in some cases also
exponentially~\cite{Armoni:2003nz}. Since $N=3$ is supposed to be
already large, these arguments may explain why the predictions of
$n$-ality have not been directly observed in numerical simulations,
which are limited in accuracy.  This situation worsens for larger $N$.

We study this issue in SU(3) gauge theory, by MC simulations.  We
measure ``wall-wall'' correlators of Polyakov lines in the
representations of rank $k=1$ (fundamental) and $k=2$ (symmetric) of
SU(3), in order to check whether their string tensions $\sigma_k$ are
consistent with $n$-ality. 

The correlators decay exponentially as $\exp (- m_k t)$, where $m_k$
is the mass of the lightest state in the corresponding representation.
For a line of size $L$, $\sigma_k$ is obtained through: $m_k =
\sigma_k L - {\pi/ (3 L)}.$ We obtained results at $\beta=5.9$ and for
two lattices: $12^3\times 24$ and $16^3\times 24$. Use of smearing and
blocking leads to a construction of new operators with a better
overlap with the lightest string state.  In Fig. 1 we show the
correlators as a function of the distance $t$ in the cases of $k=1$
and $k=2$ representations, from the runs with $L=16$; a similar figure
is provided in \cite{DPV1} for $L=12$.  The data for the correlator at
$k=1$ allow us to accurately determine the fundamental string tension,
and the two lattices give consistent results, i.e.
\Blue{$\sigma=0.0664(5)$} and \Blue{$\sigma=0.0668(3)$}\,.

\begin{figure}[t]
\centerline{\epsfig{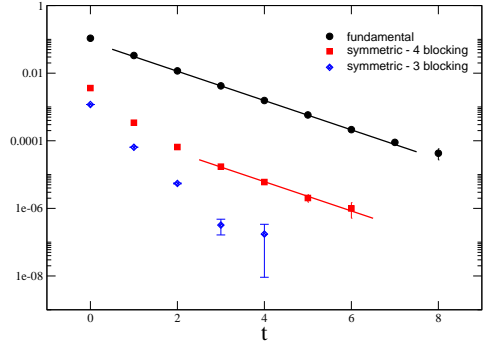}}
\vspace{-0.75truecm}
\caption {Correlator data at $L=16$, as a function of
  the distance $t$.
  The top line is an exponential fit to $k=1$
  data with $t\geq 3$, leading to an estimate of $m_1$\,; the bottom
  line is an exponential with the {\it same} exponent, which is well
  supported by the $k=2$ data for $t\geq 3$. 
 Data from the $k=2$
  representation at one less blocking step is shown in diamonds.
}
\vspace{-0.5truecm}
\end{figure}

On the other hand, such an agreement is \Blue{not} observed
for $k=2$.
However, the $L=16$ data for $k=2$
shows a clear evidence that its
asymptotic behavior is controlled by the
fundamental string; indeed, fitting the data for $t\geq 3$
we obtain $\sigma_{\rm sym}=0.070(4)$,
in agreement with $n$-ality.
Although data at small $t$, $t<3$, show a clear
contamination by heavier states, 
in the $k=2$ case the overlap with the fundamental string state
of the source operator,
obtained by performing four blocking steps after smearing, appears to be sufficient
to show the eventual asymptotic behavior.
This is not observed in the $L=16$ data using the source operator
with three blocking steps (one less) and in the $L=12$ data
(two blocking steps).
Up to the distances that we can observe, the correlators are dominated by the
propagation of a much heavier state, which would suggest 
$\sigma_{\rm sym}\approx 0.16$. Note that 
$\sigma_{\rm sym}/\sigma\approx 2.4$ is
rather close to the Casimir ratio 5/2.

In conclusion, our results provide direct
evidence that the spectrum of confining strings is according to
predictions based on $n$-ality. Torelon correlations in the rank-2
symmetric channel appear to be well reproduced  by a two-exponential
picture, in which the heavier string state has $\sigma_2/\sigma_1 \sim
C_{\rm sym}/C_{\rm f} = 5/2$, and the torelon has a much smaller
overlap with the lighter $k$-string.

\section{Topological susceptibility across $T_c$}

At $T=0$ the anomalous breaking of the axial $U(1)$ flavour symmetry
explains the heavier singlet state $\eta'$, whose mass is related to
the pure gauge topological susceptibility $\chi$ through the
Witten-Veneziano (WV) formula $m^2_{\eta^\prime}= 4 N_f \chi /
f^2_\pi$, up to $\mathcal O(1/N)$ corrections.
In order to clarify the pattern of chiral symmetry breaking, we extend
our study of the dependence on the $\theta$ angle in SU($N$) gauge
theories~\cite{DelDebbio:2002xa} to the case of finite temperature.

At $T>0$, chiral symmetry is restored at a critical temperature,
$T_{\mathrm{ch}}$. The nature of this phase transition is relevant to
understanding the behaviour of hadronic matter under extreme
conditions. Below $T_{\mathrm{ch}}$, we expect the WV formula to
hold. In order to test the WV mechanism at finite $T$, we study $\chi$
as $T_{\mathrm{ch}}$ is approached from below.  Above
$T_{\mathrm{ch}}$ the picture is rather different. Here instanton
calculus gives contributions for $\chi$ that are exponentially
suppressed, $\chi\sim \exp\left(-N \right)$. Since the transition
between high- and low-$T$ is not fully understood, we also present
results above $T_{\mathrm{ch}}$.

Our study is carried out for $N=4,6$, on asymmetric lattices
with different time extensions $L_t=6,8$ and constant aspect ratio
$L_t/L_s=1/4$. 
The physical temperature is a function of
lattice spacing and $L_t$,
$T=1/a(\gamma) L_t$, where $\gamma=\beta/2 N^2$. For each
$L_t$, $\gamma$ is chosen around $\gamma_c$ which corresponds to the
critical temperature $T_{\mathrm{ch}}$\,. 
We define absolute and reduced temperatures through:
\begin{eqnarray}
&&T(L_t,\gamma) 
\equiv {T / \sqrt{\sigma}} = {1/(L_t \sqrt{\sigma(\gamma)})},\\
&& 
t(L_t,\gamma) \equiv {T(L_t,\gamma) /
  T(L_t,\gamma_c(L_t))} - 1\, ,
\label{redtemp}
\end{eqnarray}
($\sigma$ is computed on symmetric lattices, $T=0$). 

$\chi$ is defined using cooling; we compute: 
\begin{equation}
R(L_t,\gamma)  \equiv 
{\chi_t(L_t,\gamma) / \chi_t(\infty,\gamma)},     
\end{equation}
where $\chi_t(\infty,\gamma)$ is computed at $T=0$.
\begin{figure}[t]
\centerline{\psfig{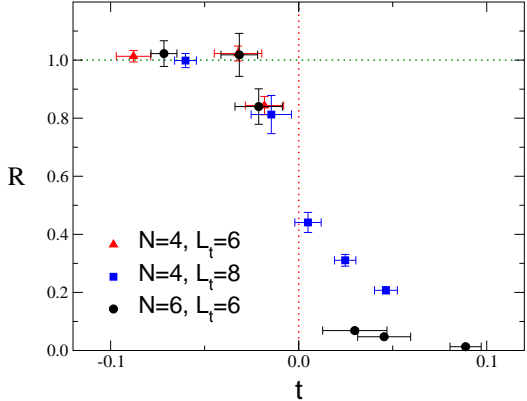}}
\vspace{-0.75truecm}
\caption{$R$ as a function of reduced temperature.}
\vspace{-0.5truecm}
\end{figure}

Data for the scaling ratio $R$ are displayed in Fig. 2.  Its behavior
is drastically different in the low- and high-$T$ phases.  At low-$T$,
all data for $N=4$, $L_t=6,8$ and $N=6$, $L_t=6$ appear to lie on the
same curve, showing both that scaling corrections are small and the
large-$N$ limit is approached fast.  $R$ remains constant, compatible
with $1.0$. Only close to $T_{\mathrm{ch}}$, i.e. for $T\simeq 0.98
T_{\mathrm{ch}}$\,, it appears to decrease.  Thus the physical picture
does not change as long as the system remains in the confined phase.
On the other hand, above the transition, we observe a clear
suppression of $\chi_t$, which appears much stronger at $N=6$.

Thus, in the \Blue{confined} phase, topological properties remain
substantially unchanged up to $T_{\mathrm{ch}}$. In the
\Blue{high-$T$} phase, there is a sharp change of regime where $\chi$
is largely suppressed. Similar results have been recently presented
in~\cite{Lucini:2004yh}. The MC data suggests that such a drop grows
with increasing $N$.  This suppression at large $N$ supports the
hypothesis that topological properties in the high-$T$ phase are
essentially determined by instantons~\cite{Kharzeev:1998kz}.

\end{document}